\documentclass[prd,twocolumn,floatfix,preprintnumbers,showpacs]{revtex4}
\usepackage{graphicx}
\usepackage{dcolumn}
\usepackage{bm}
\usepackage{color}

\def\lsim{\mathrel{\mathop
  {\hbox{\lower0.5ex\hbox{$\sim$}\kern-0.8em\lower-0.7ex\hbox{$<$}}}}}
\def\gsim{\mathrel{\mathop
  {\hbox{\lower0.5ex\hbox{$\sim$}\kern-0.8em\lower-0.7ex\hbox{$>$}}}}}

\begin{document}

\newcommand{\half}{{1\over2}}
\newcommand{\bk}{{\bf k}}
\newcommand{\Ocdm}{\Omega_{\rm cdm}}
\newcommand{\ocdm}{\omega_{\rm cdm}}
\newcommand{\OM}{\Omega_{\rm M}}
\newcommand{\OB}{\Omega_{\rm B}}
\newcommand{\oB}{\omega_{\rm B}}
\newcommand{\OX}{\Omega_{\rm X}}
\newcommand{\cltt}{C_l^{\rm TT}}
\newcommand{\clte}{C_l^{\rm TE}}
\newcommand{\mwdm}{m_{\rm WDM}}
\newcommand{\mnu}{\sum m_{\rm \nu}}

\input epsf

\preprint{DFPD/04/A26, LAPTH-1081/04}
\title{CMB lensing extraction and primordial non-Gaussianity}
\author{Julien Lesgourgues,$^{1,2}$ Michele Liguori,$^{3,2}$
Sabino Matarrese,$^{3,2}$ Antonio Riotto,$^2$}
\affiliation{
$^1$Laboratoire de Physique Th\'eorique LAPTH, F-74941
Annecy-le-Vieux Cedex, France\\
$^2$ INFN, Sezione di Padova,
Via Marzolo 8, I-35131 Padova, Italy\\
$^3$ Dipartimento di Fisica ``G. Galilei'', Universit\`a di Padova,
Via Marzolo 8, I-35131 Padova, Italy
}
\date{December 20, 2004}
\begin{abstract}
The next generation of CMB experiments should get a better handle on
cosmological parameters by mapping the weak lensing deflection field,
which is separable from primary anisotropies thanks to the
non-Gaussianity induced by lensing. However, the generation of
perturbations in the Early Universe also produces a level of
non-Gaussianity which is known to be small, but can contribute to the
anisotropy trispectrum at the same level as lensing. In this work, we
study whether the primordial non-Gaussianity can mask the lensing
statistics. We concentrate only on the ``temperature quadratic
estimator'' of lensing, which will be nearly optimal for the Planck
satellite, and work in the flat-sky approximation. We find that
primordial non-Gaussianity contaminates the deflection field estimator
by roughly $(0.1 f_{NL})$\% at large angular scales, which represents
at most a 10\% contribution, not sufficient to threaten lensing
extraction, but enough to be taken into account.
\end{abstract}

\maketitle

{\it Introduction --}
Cosmic Microwave Background (CMB) anisotropies are of considerable interest
for cosmology because after cleaning the observed temperature and
polarization maps from various foregrounds, one obtains a picture of
cosmological perturbations on our last-scattering surface. The power
spectra of primary anisotropies are related to various 
cosmological parameters, and depend on the physical evolution
mainly before the time of decoupling (and also, more weakly, on its
recent evolution, through the integrated Sachs-Wolfe effect and the
angular diameter--redshift relation).

It was realized recently that CMB anisotropies encode even
more cosmological information than expected, because it should be
possible in a near future to measure the deflection field caused by
the weak lensing of CMB photons by the large scale structure of the
neighboring universe at typical redshifts $z\sim 3$ \cite{Ber97,LE}. The
power spectrum of the deflection field encodes some information
concerning structure formation mainly in the linear or quasi-linear
regime, and is therefore extremely useful for measuring parameters
like the total neutrino mass or the dark energy equation-of-state,
which mildly affect the primary anisotropy \cite{Kap03}. 
So, the next generation of
CMB experiments could output for free a Large Scale Structure (LSS)
power spectrum, without suffering like galaxy redshift surveys from
the systematics induced by mass-to-light bias and by strong
non-linear corrections on small scales at $z\leq0.2$.

There are several methods on the market for extracting the deflection
map \cite{QE1,QE2,QE3,IE}, 
all based on the non-Gaussianity induced by lensing
\cite{Ber97}. These
methods start from the assumption that both the primary anisotropies
and the deflection field are Gaussian; they also
assume that the noise present in the temperature and polarization maps
is Gaussian and uncorrelated with the signal.

None of these assumptions is exactly true. Amblard et al. \cite{Amb04}
already estimated to which extent the lensing extraction will be
biased, first, by the non-Gaussianity of the lensing potential caused by the
non-linear growth of matter perturbations on small scales, and second,
by the imperfect cleaning of the
CMB maps from the kinetic Sunyaev-Zel'dovich effect, which also has a
blackbody spectrum, induces non-Gaussianity, and features spatial
correlations with many of the structures responsible for the
lensing. Both effects were found to be relevant (i.e., to induce a
significant bias in the estimators). However, they are small enough to
preserve the validity of the method.

The purpose of this work is to relax the first of the previous
assumptions, and to consider realistic situations, in which one
cannot avoid a small level of non-Gaussianity to be produced in the
Early Universe. Non-Gaussianity emerges 
as a key observable to discriminate among competing scenarios for the 
generation of cosmological perturbations and is one of the primary targets of 
present and future CMB satellite missions \cite{NG}.
Indeed, despite the simplicity of the inflationary paradigm \cite{reviewlr}, 
the mechanism by which  cosmological curvature 
perturbations are generated  is not
yet established. In the standard slow-roll scenario associated
to one-single field models of inflation, the observed density 
perturbations are due to fluctuations of the inflaton field itself when it
slowly rolls down along its potential.  
In the curvaton 
mechanism~\cite{curvaton} the final curvature perturbation  is 
produced from an initial isocurvature perturbation associated with the
quantum fluctuations of a light scalar field (other than the inflaton), 
the curvaton, whose energy density is negligible during inflation. 
Recently, other  mechanisms for the generation of cosmological
perturbations have been proposed, the inhomogeneous reheating scenario 
\cite{gamma1}, the ghost inflationary scenario \cite{ghost}, 
and the D-cceleration scenario~\cite{dacc}, just to mention 
a few. Single-field slow-roll inflationary model
inflation itself produces a negligible amount of 
non-Gaussianity, and the dominant contribution comes from the evolution of 
the ubiquitous second-order perturbations after inflation. However, 
alternative models for the generation of
perturbations might produce much stronger primordial non-Gaussianity. 
Therefore, non-Gaussianity in the CMB maps  from
primordial fluctuations could mask the non-Gaussianity from lensing
distortions.  We will see later that if
one expands the power spectrum of the lensing estimator in powers of
the gravitational potential $\Phi \sim 10^{-5}$, the contribution from
primordial non-Gaussianity appears at the same order as the lensing
power spectrum itself. Therefore, a precise computation is needed in
order to understand whether the primordial non-Gaussianity could
affect lensing extraction.

\vspace{0.5cm}

{\it Lensing extraction with quadratic estimators --} Weak lensing
induces a deflection field ${\bf d}$, i.e., a mapping between the
direction of a given point on the last scattering surface and the
direction in which we observe it. At leading order \cite{IE}
this deflection field can be written as the gradient a lensing potential, 
${\bf d}=\nabla \phi$.

In the limit of Gaussian primordial fluctuations, the unlensed
anisotropies obey Gaussian statistics, and in the flat-sky
approximation their two-dimensional Fourier modes are fully described
by the power spectra $\tilde{C}^{ab}_l$ where $a$ and $b$ belong to
the $\{T, E, B\}$ basis. Weak lensing correlates the lensed multipoles
\cite{Sel96,Ber97} according to
\begin{equation}
\langle a({\bf l}) b({\bf l'}) \rangle_{\rm CMB}
= (2 \pi)^2 \delta({\bf l}+{\bf l'}) \tilde{C}^{ab}_{l}
+ f^{ab}({\bf l}, {\bf l'}) \phi({\bf l}+{\bf l'})
\label{ab}
\end{equation}
where the average holds over different realizations (or different
Hubble patches) of a given cosmological model with fixed primordial
spectrum and background evolution (i.e. fixed cosmological
parameters). In this average, the lensing potential is also kept fixed
by convention, which makes sense because the CMB anisotropies and LSS
that we observe in our past light-cone are statistically independent,
at least as long as we neglect the integrated Sachs-Wolfe effect. The
above function $f^{ab}$ is defined in \cite{QE2} and takes a simple form
in the case $ab=TT$:
\begin{equation}
f^{TT}({\bf l}, {\bf l'})
= C^{TT}_{l}
({\bf l}+{\bf l'}) \cdot {\bf l}
+
C^{TT}_{l'}
({\bf l}+{\bf l'}) \cdot {\bf l}'~.
\end{equation}
Our study will be based on the quadratic estimator method
of Hu \& Okamoto \cite{QE1,QE2,QE3} (which is
equivalent in terms of precision to the alternative
iterative estimator method of Hirata \& Seljak 
\cite{IE} as long as
CMB experiments will make noise-dominated measurements of the B-mode,
i.e., at least for the next decade).
By inverting Eq.~(\ref{ab}), one builds a quadratic combination of 
the temperature and polarization observed Fourier modes
\begin{equation}
{\bf d}^{ab}({\bf L}) =
\frac{i {\bf L} A^{ab}_L}{L^2}
\int \!\!\! \frac{d^2 {\bf l}_1}{(2 \pi)^2}
a({\bf l_1}) b({\bf l_2}) g^{ab}({\bf l_1},{\bf l_2})
\label{defd}
\end{equation}
where ${\bf l_2} = {\bf L} - {\bf l_1}$, and in which
the normalization condition
\begin{equation}
A^{ab}_L = L^2 \left[ \frac{d^2 {\bf l}_1}{(2 \pi)^2}
f_{TT}({\bf l}_1,{\bf l}_2) g^{ab}({\bf l}_1,{\bf l}_2)
\right]^{-1}
\end{equation}
ensures that ${\bf d}^{ab}$ is 
an unbiased estimator of the lensing potential:
\begin{equation}
\langle {\bf d}^{ab}({\bf L})\rangle_{\rm CMB}
= i {\bf L} \phi({\bf L}) = {\bf d}({\bf L})~.
\label{avd}
\end{equation}
Note that, so far, the coefficients $g^{ab}({\bf l}_1,{\bf l}_2)$ are
still arbitrary. From the observed temperature and polarization maps,
one could compute each mode of ${\bf d}^{ab}$ and obtain various
estimates of the deflection modes, precise up to cosmic
variance and experimental errors. In order to quantify the total
error, it is necessary to compute the power spectra of the quadratic
estimators
\begin{equation}
\langle{\bf d}^{ab*}({\bf L}){\bf d}^{ab}({\bf L})\rangle =
(2\pi)^2 \delta({\bf L}-{\bf L}') C_L^{dd(ab)}
\end{equation} 
where the average is now taken over both CMB and LSS realizations,
since $\phi({\bf L})$ is also a stochastic quantity. In this
definition, the power spectra are written with a superscript {\small
$dd(ab)$} in order to be distinguished from the actual power spectrum
of the true deflection field.  These spectra feature the four-point
correlation function of the observed (lensed) Fourier modes $\langle
a({\bf l}_1) b({\bf l}_2) a({\bf l}_3) b({\bf l}_4)\rangle$, which
should be expanded at order two in $\phi({\bf L})$ in order to catch
the leading non-Gaussian contribution.

The four-point correlation functions are composed as usual of a
connected and an unconnected piece. The connected piece is by
definition a function of the power spectra $C_{l}^{ab}$ in which we
now include all sources of variance: cosmic variance, lensing
contribution and experimental noise. The unconnected piece is a
function of the same spectra plus the deflection spectrum
$C_{l}^{dd}$, and as usual it can be decomposed in three terms
corresponding to the different pairings of the four indices
\cite{TRI1}: $(l_1,l_2)$, $(l_3,l_4)$ or $(l_1,l_3)$, $(l_2,l_4)$ or
$(l_1,l_4)$, $(l_3,l_2)$.  The first term leads to considerable
simplifications when it is plugged into the expression of the
quadratic estimator power spectrum, and the result is simply
$C_l^{dd}$, as one would expect naively from squaring Eq.~(\ref{avd}).
The other terms lead to more complicated expressions that we will
write as two noise terms:
\begin{equation}
C_L^{dd(ab)} =
C_L^{dd} + [N_{\rm c}]^{ab}_{L} + N^{ab}_{L}~,
\label{clddab}
\end{equation}
which represent respectively the contribution
from the connected piece and from the two non-trivial terms of
the unconnected piece \cite{TRI1,Coo02}
(later, we will give the exact expressions in the case $aa=TT$).  
In order to get an efficient
estimator, we should adopt the set of coefficients $g^{ab}({\bf l}_1,{\bf
l}_2)$ which minimize the noise terms.  It is actually much easier to
minimize the connected term only, which leads to the simple results
\begin{equation}
g^{aa}({\bf l}_1,{\bf l}_2)
= 
\frac{f_{aa}({\bf l},{\bf l}')}{2 C_{l}^{aa} C_{l'}^{aa}} 
\quad {\rm and} \quad
[N_{\rm c}]^{aa}_{L} = A^{aa}_L
\end{equation}
for $a\!=\!b$ (for $a \!\neq\! b$ see \cite{QE2}).
With such a choice, the unconnected piece contribution
$N^{ab}_{L}$ can be shown to be smaller than $A^{aa}_L$, 
but not completely negligible \cite{Coo02}.

The various estimators ${\bf d}^{ab}$ can be constructed for each pair of
modes, except for the pair $BB$, because the spectrum $C_l^{BB}$ is
dominated by lensing at least on small scales, which invalidates the
present method. So, the quadratic estimator technique would not be
optimal for long-term CMB experiments with cosmic-variance-dominated
measurement of the $B$ mode \cite{IE,Smi04}. For an experiment of given
sensitivity, the five other estimators can be combined into a final
minimum variance estimator, which gives the best possible
estimate of the deflection field by weighing each estimator
accordingly to its noise level. The sensitivity of the Planck 
satellite \cite{Planck} is slightly 
above the threshold for successful lensing extraction, but only at
intermediate angular scales, and with essentially all the signal coming
from the ${\bf d}^{TT}$ estimator. The following generation of experiments
-- such as the CMBpol or Inflation probe project \cite{CMBpol} --
should obtain the lowest
noise level from the ${\bf d}^{EB}$ estimator \cite{QE2}.

We summarized here the quadratic estimator method, which assumes that
both the primary anisotropies and the lensing potential are Gaussian.
We will now study the impact of primordial non-Gaussianity.  This
program is numerically cumbersome, and we will only concentrate on the
${\bf d}^{TT}$ estimator, which is the only relevant one for Planck,
while sticking to the flat-sky approximation in which numerical
computations are much quicker.

\vspace{0.5cm}

{\it Contributions from primordial non-Gaussianity --}  The
two-dimensional Fourier modes of both temperature anisotropies and the
lensing potential can be related to the stochastic three-dimensional
modes of the primordial gravitational potential $\Phi({\bf k})$,
multiplied by a transfer function which accounts for its
time-evolution. The fact of writing a unique stochastic function can
bring some confusion, because the modes $\Phi({\bf k})$ which appear
in the CMB and lensing expressions represent fluctuations at very
different redshifts: the first ones before decoupling, the second at
$z\sim3$, i.e. in the neighboring universe. So, as long as we neglect the
integrated Sachs-Wolfe effect, it is convenient to introduce two
statistically independent functions $\Phi^{\rm CMB}({\bf k})$ and
$\Phi^{\rm LSS}({\bf k})$, sharing the same statistical properties,
but sourcing respectively $a({\bf l})$ and $\phi({\bf l})$.

The true non-Gaussian potential $\Phi^{\rm X}_{NL}$ ({\sc x} = {\sc
cmb} or {\sc lss}) can be expanded in real space in powers of a
Gaussian potential $\Phi^{\rm X}_{L}$.  In Fourier space and at order
three,
\begin{equation}
\Phi^X_{NL}({\bf k})=\Phi^X_{L}({\bf k})+\Phi^X_A({\bf k})+\Phi^X_B({\bf k})
\label{expansion}
\end{equation}
with
\begin{eqnarray}
\Phi^X_A({\bf k})\!\!\! &=& \!\!\! f_{NL} \! \left[
\int \!\! \frac{d^3 {\bf p}}{(2 \pi)^3} 
\Phi^X_{L}({\bf k\!+\!p}) \Phi^{X*}_{L}({\bf p})\!
- \!(2\pi)^3 \delta({\bf k}) \overline{\Phi^2_L} \right]
\nonumber\\
\Phi^X_B\!({\bf k}) \!\!\! &=& \!\!\! g_{NL} \!\!
\int \!\!\! \frac{d^3 {\bf p}_1}{(2 \pi)^3} \! \frac{d^3 {\bf p}_2}{(2 \pi)^3} 
\Phi^{X*}_{L}\!({\bf p_1}) \Phi^{X*}_{L}\!({\bf p_2})
\Phi^X_{L}\!({\bf p_1\!\!+\!\!p_2\!\!+\!\!k})
\nonumber
\\
{\rm and}~ &&
\overline {\Phi^2_L} = \int \frac{d^3 {\bf k}}{(2 \pi)^3} P_{\Phi}(k)~.
\nonumber
\end{eqnarray}
We have parametrized the primordial non-Gaussianity by a quadratic
and a cubic term in the gravitational potential. They are proportional
to the dimensionless parameters $f_{\rm NL}$ and $g_{\rm NL}$, respectively. 
The theoretically predicted
parameter $f_{\rm NL}$ appears as a kernel in Fourier space, rather than a
constant, in most of the scenarios for the generation of the
cosmological perturbations, while theoretical predictions for the
parameter $g_{\rm NL}$ are still lacking 
\footnote{The parameter $g_{\rm NL}$
could be rather sizeable in some scenarios for the generation of the
cosmological perturbations. For instance, in the curvaton scenario
\cite{curvaton} $g_{\rm NL}$ can be as large as $f_{\rm NL}^2\sim 10^4$.}.
This gives rise to an angular modulation of the quadratic
non-linearity, which might be used to search for specific signatures
of inflationary non-Gaussianity in the CMB. In this paper, however,
we restrict ourselves to the simplest case and assume $f_{\rm NL}$ and
$g_{\rm NL}$ as mere phenomenological multiplicative constants. Under this
assumption, the {\sl WMAP} team has measured the bispectrum to obtain
the tightest limit to date, $-58<f_{\rm NL}<134$ ($95\%$) \cite{k}.  On
the other hand, no observational bound has been set on $g_{\rm NL}$ from
the observed trispectrum. However, one can simply notice from 
Eq.~(\ref{expansion}) that the small parameter $(f_{\rm
NL}\Phi)$ contributes at the same order as $(\sqrt{g_{\rm NL}}~\Phi)$: so, by
comparing with the $f_{\rm NL}$ bound, it is likely that values of
order $\sqrt{g_{\rm NL}} \sim 100$ are still allowed by the data.

As far as lower bounds are concerned, one should keep in mind that
although single-field slow-roll inflation itself produces a
negligible amount of non-Gaussianity, the dominant contribution
comes from the evolution of the ubiquitous second-order perturbations
after inflation.  This effect {\it must exist} regardless of the
inflationary model, setting the minimum level of non-Gaussianity in
the cosmological perturbations at order $f_{\rm NL}\!\sim\! g_{NL} \!\sim\! 1$.

\begin{figure*}[ht!]
\begin{center}
\vspace{1cm}
\includegraphics[angle=0,width=9cm]{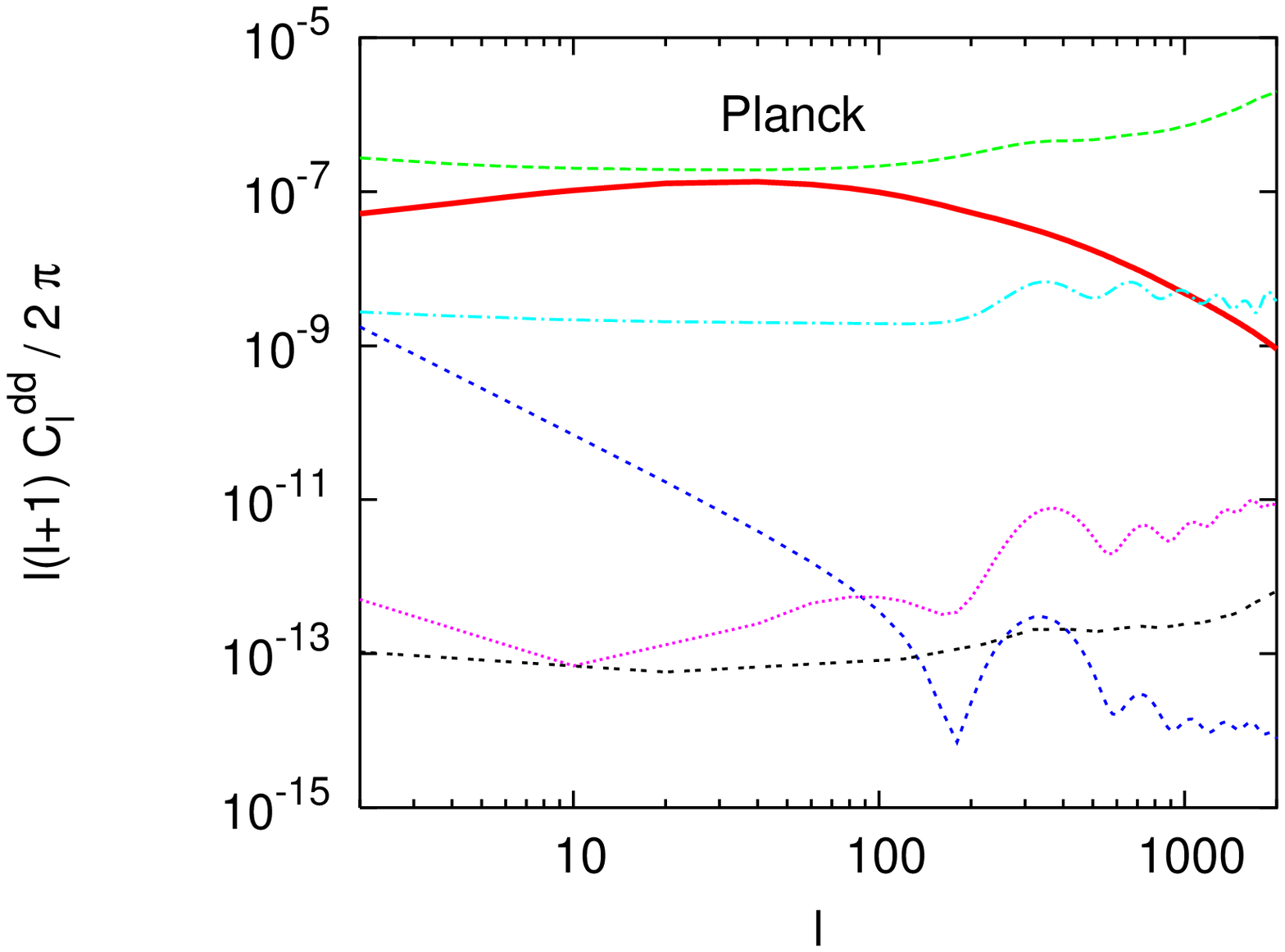}
\hspace{-1cm}
\includegraphics[angle=0,width=9cm]{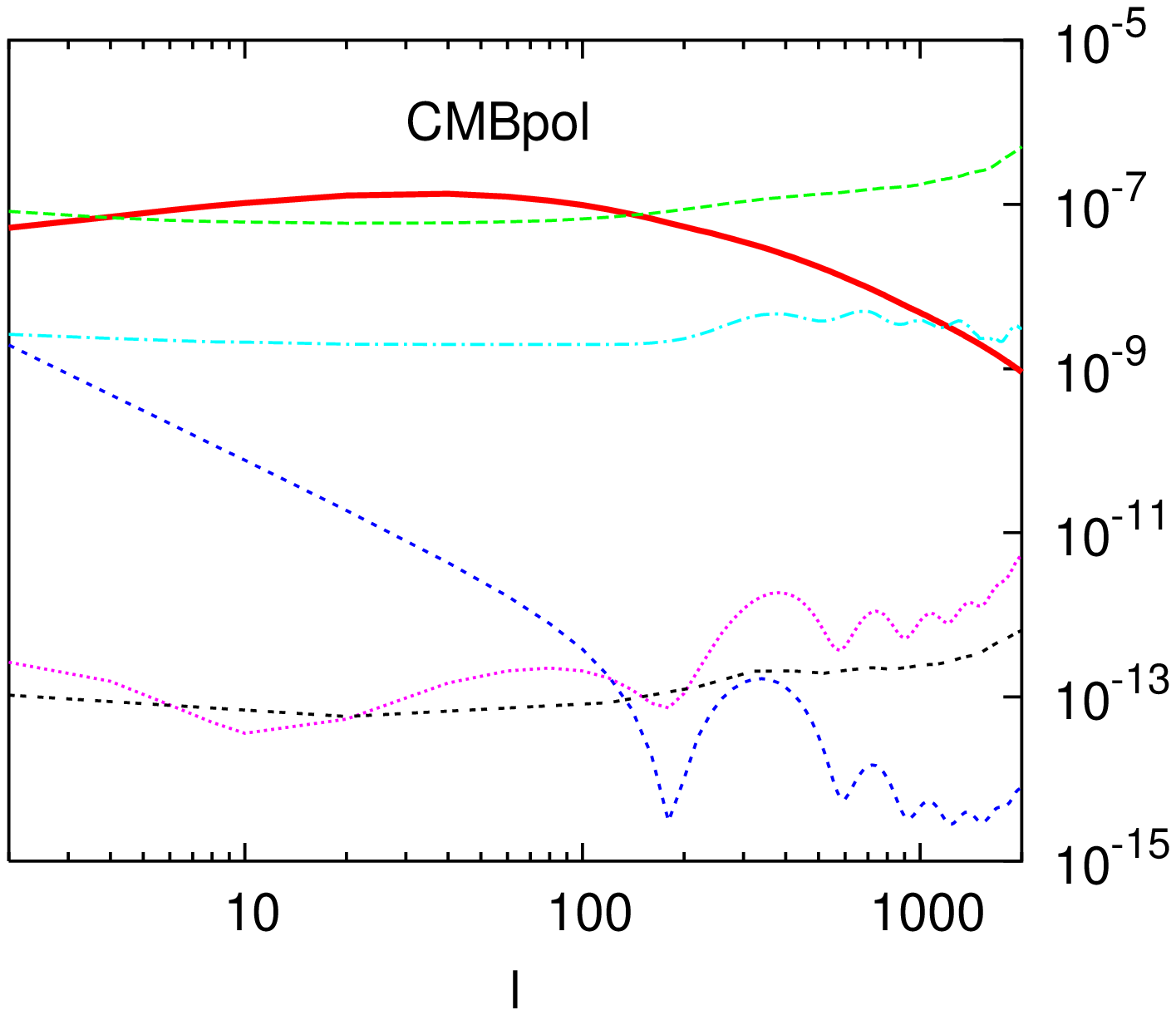}
\end{center}
\vspace{-0.5cm}
\caption{\label{fig} Various contributions to the variance of a single
mode of the estimator ${\bf d}^{TT}({\bf l})$, for the case of Planck
(left) and CMBpol (right).  The thick (red) curve shows the variance
of the signal $C_l^{dd}$.  Other curves represent (from top to bottom
at $l\!\sim\!2000$) the noise variance from the connected part of the
four-point correlation function, from the lensing trispectrum, and
from the terms $B$, $A2$ and $A1$ in the primordial non-Gaussianity
trispectrum (displayed for $f_{NL}\!=\!g_{NL}^{1/2}\!=\!100$).}
\end{figure*}
In order to evaluate the impact of these extra contributions on the
power spectrum $C_L^{dd(a,b)}$, we should first recompute the
four-point functions $\langle a_{l_1}^{m_1} b_{l_2}^{m_2} a_{l_3}^{m_3}
b_{l_4}^{m_4}\rangle$, working as before at order six in the gravitational
potential. Non-zero contributions can arise only from terms with an
even number of $\Phi^{\rm CMB}_L({\bf k})$ and $\Phi^{\rm LSS}_L({\bf
k})$ factors. The standard calculation of the previous section
included terms in which either two multipoles were lensed at order one
in $\phi({\bf l})$, or one multipole was lensed at order two (the
later terms contributes only to the connected piece). In addition, we
should now consider terms in which:
\begin{itemize}
\item[A.] the four multipoles are unlensed, but two of them include the term
$\Phi^{\rm CMB}_A$,
\item[B.] the four multipoles are unlensed, one of them includes the term
$\Phi^{\rm CMB}_B$,
\item[C.] one of the four multipoles is lensed at order one in $\phi({\bf l})$,
which includes the term
$\phi^{\rm LSS}_A$.
\end{itemize} 
\mbox{ }\\

The last term C vanishes because $\phi^{\rm LSS}_A$ has zero average.
So, at leading order, lensing and primordial non-Gaussianity effects
are completely separable, and we simply need to add corrections from
the primordial non-Gaussianity trispectrum, which is given in 
Okamoto \& Hu \cite{TRI2} (who computed it in the Sachs-Wolfe
approximation).
First, one needs to compute the integrals
\begin{eqnarray}
F_l(r_1,r_2) &=& \frac{2}{\pi}
\int k^2 dk \, P_{\Phi}(k) j_l(k r_1) j_l(k r_2)~, 
\\
\alpha_{l}(r) &=& \frac{2}{\pi}
\int k^2 dk \, \Delta_{l}(k) j_l(k r)~,
\\
\beta_{l}(r) &=& \frac{2}{\pi}
\int k^2 dk \, P_{\Phi}(k) \Delta_{l}(k) j_l(k r)~,
\end{eqnarray}
where $\Delta_{l}(k)$ is the radiation transfer function for the temperature,
normalized to $\Phi\!=\!1$ in the early universe. We checked that our
functions $\alpha_{l}(r)$ and $\beta_{l}(r)$ (computed with a slightmy modified
version of {\sc cmbfast} \cite{CMBFAST}) perfectly agree with
those of \cite{Kom01}.
Each of the $A$ and $B$-type trispectra are composed of three parts, 
which can be simply expressed in terms of the intermediate quantities
\begin{eqnarray}
&(P_A)^{{l_1} {l_2}}_ {{l_3} {l_4}}&\!\!\!(L)
=  
4 f_1^2 \!\!\!
\int \!\! r_1^2 d r_1
\!\!
\int \!\! r_2^2 d r_2
\,
F_L(r_1,r_2) \times
\nonumber \\
&&
\,\,\,\,\,\,
\left[\alpha_{l_1}\!(r_1) \beta_{l_2}\!(r_1)
+{\rm c.p.}\right]
\left[\alpha_{l_3}\!(r_2) \beta_{l_4}\!(r_2)
+{\rm c.p.}\right]~,
\nonumber \\
&
(P_B)^{{l_1} {l_2}}_ {{l_3} {l_4}}&
\!\! =
2 f_2 \!\!
\int \!\! r^2 d r
\left[
\alpha_{l_1}\!(r) \beta_{l_2}\!(r) \beta_{l_3}\!(r) \beta_{l_4}\!(r)
+
{\rm c.p.}\right]~,
\nonumber
\end{eqnarray}
where c.p. means circular permutation of the indices $l_i$.
The final expression for the power spectrum of the quadratic estimator
for any mode ${\bf L}$ of modulus $L$,
including the three trispectra induced by lensing and primordial
non-Gaussianity, reads
\begin{eqnarray}
&& C_L^{dd(TT)} \! = 
C_L^{dd}  +  A_L^{TT} 
\nonumber
\\
&&+ \frac{{A^{TT}_L}^2}{L^2}
\!\! \int \!\!\! \frac{d^2 {\bf l}_1}{(2 \pi)^2}
\! \frac{d^2 {\bf l}'_1}{(2 \pi)^2}
g^{TT}\!({\bf l}_1,\!{\bf l}_2)
g^{TT}\!({\bf l}'_1,\!{\bf l}'_2) \times
\nonumber
\\
&&\left[
2 \, |{\bf l}_1\!-{\bf l}'_1|^{-2}
C_{|{\bf l}_1-{\bf l}'_1|}^{dd}
f^{TT}({\bf l}_1,\!-{\bf l}'_1)
f^{TT}({\bf l}_2,\!-{\bf l}'_2)
\right.
\nonumber
\\
&&\left.
+
(P_A)^{{l_1} {l_2}}_{{l'_1} {l'_2}}(L)
+
2 (P_A)^{{l_1} {l'_1}}_{{l_2} {l'_2}}(|{\bf l}_1-{\bf l}'_1|)
+
3 (P_B)^{{l_1} {l_2}}_{{l'_1} {l'_2}}
\right]~~~~~~~~
\label{fr}
\end{eqnarray}
with ${\bf l}_2={\bf L} - {\bf l}_1$ and
${\bf l}'_2={\bf L} - {\bf l}'_1$.
The variance of the signal $C_L^{dd}$ comes from the first term of the
lensing trispectrum, while the noise variance $A_L^{TT}$ comes from
the connected part of the four-point correlation function.
We can identify all the other sources of noise inside the integral.
The third line of Eq.~(\ref{fr}) contains the two other terms of the
lensing trispectrum, which give equal contributions (as can be shown
after a change of variable). 
The last line contains
the three terms of the $A$-type trispectrum (the last two contribute equally) 
and the three terms of the $B$-type trispectrum (three equal contributions).
Let us call $A1$, $A2$ and $B$ the terms of the last line. 

The numerical integration of Eq.~(\ref{fr}) could be exceedingly long,
due to the four-dimensional integral in Fourier space (which converges
only when the upper bound of integration is taken around the
Silk-damping cut-off, $l_{\rm max} \sim 2500$) and of the
two-dimensional integral in $r$ space. Fortunately, the second
integral can be limited to values of $r$ close to the comoving
distance to the last-scattering surface, far from which $r^2
\alpha_l(r)$ is negligible. 
Here, we will integrate $r=\tau_0-\tau$
from $0.03 \tau_*$ to $1.3 \tau_*$
with a conservative step $d\tau=0.01 \tau_*$ (where $\tau$ stands for
conformal time, $\tau_0$ is evaluated today, and
$\tau_*$ is the conformal time at recombination,
defined as the peak of the visibility function).

The integration in $l$-space can be sped up by taking into account
various symmetries, and also by noting that for the terms $A1$ and $B$
the integral over ${\bf l}_1$ and ${\bf l}_1'$ are separable. Then,
the integration effectively reduces to a two-dimensional one. For
these terms, we integrate first on ${\bf l}_1$, then on $r_1$ and
$r_2$, and obtain the full result in few minutes.  The term $A2$ and
the lensing term require a real four-dimensional integral and more CPU
time. Fortunately, the quantities to integrate are very smooth in
$l$-space, and one can considerably reduce the computing time by
choosing a large step without loosing precision. We checked that
$\Delta l\sim 30$ is by far sufficient.

\vspace{0.5cm}

{\it Results and conclusions --} We take a fiducial $\Lambda$CDM model
with $\Omega_b\!=\!0.05$, $\Omega_c\!=\!0.25$,
$\Omega_{\Lambda}\!=\!0.70$, $h\!=\!0.65$, no reionization and a
scale-invariant primordial spectrum $P_{\Phi}(k)\!=\!6.204 \times
10^{-11} k^{-3}$. For instrumental noise, we consider the cases of
Planck HFI (three channels) and of the CMBpol project, with a
sensitivity described by the same parameters as in \cite{LPP}. We show
in Fig.~\ref{fig} the various contributions to the estimator power
spectrum, as computed from Eq.~(\ref{fr}). Note that we are plotting
the variance of a single mode ${\bf l}$, and not the error on the
reconstructed deflection power spectrum, which can be lowered by
combining all modes of given wavenumber $l$ and by binning the data
(this is why Planck is likely to make a reasonable detection of the
deflection power spectrum at intermediate $l$'s \cite{QE2}, although
the noise variance is slightly larger than the signal variance in
Fig.~\ref{fig}). The contributions from the primordial non-Gaussianity
terms $A$ and $B$ scale respectively like $f_{NL}^2$ and $g_{NL}$, and
here they are shown for $f_{NL}^{\phantom
1}\!=\!\sqrt{g_{NL}}\!=\!100$.

The contamination from primordial non-Gaussianity appears to arise
mainly at low $l$, from the $A1$-type term. On these scales 
it would be necessary to perform an exact all-sky
computation in order to make a precise prediction. However, the error
caused by the flat-sky approximation even at low $l$ is usually small
\cite{TRI1}.

The noise induced by primordial non-Gaussianity is responsible for
roughly $(0.1 f_{NL})$\% of the amplitude of the estimator ${\bf d}^{TT}({\bf
l})$ in the range $2<l<10$: so, around 10\% for the largest possible value
of $f_{NL}$, and around 0.1\% for standard slow-roll inflationary models.
In the range $100<l<1000$, the
contribution is roughly of order $(10^{-3} f_{NL})$\% from the 
$A2$ term, and $(0.01 \sqrt{g_{NL}})$\% from the $B$ term.  


If in the near future $f_{NL}$ appears to be large, it will be
measured independently using the three-point correlation function. It
should then be possible to subtract to some extent the bias induced by
primordial non-Gaussianity when reconstructing the power spectrum
$C_L^{dd}$, as suggested in \cite{Coo03,Amb04} for other
sources of bias.

We conclude that primordial non-Gaussianity should be taken into account
if it is as large as to saturate the present upper bounds, but that in no case 
it will represent a dangerous issue for lensing extraction.

\section*{Acknowledgments}
We would like to thank W.~Hu and E.~Komatsu for useful exchanges.
This work was carried during a six-month visit of J.~L. at the University of
Padova, supported by INFN and by the Dipartimento di Fisica Galileo
Galilei.

\end{document}